\documentclass[aps,12pt,superscriptaddress,preprintnumbers,
                secnumarabic,nofootinbib,showpacs]{revtex4}

\usepackage{epsfig}
\usepackage{bbm}
\usepackage{amssymb}

\newcommand{\lbfig}[1]{\refstepcounter{fig} \label{#1} }
\newcounter{fig}

\begin{document}

\preprint{NIKHEF-2007-09, SPIN-07-11, ITP-UU-07-18}

\title{\Large Quantum radiative corrections to slow-roll
inflation}

\author{Ante Biland\v{z}i\'{c}}
\email[]{A.Bilandzic@students.uu.nl, anteb@nikhef.nl}
\affiliation{Institute for Theoretical Physics (ITP) \& Spinoza
Institute,
             Utrecht University, Leuvenlaan 4, Postbus 80.195,
              3508 TD Utrecht, The Netherlands}
\affiliation{NIKHEF, Kruislaan 409, 1098 SJ Amsterdam, The
Netherlands}

\author{Tomislav Prokopec}
\email[]{T.Prokopec@phys.uu.nl}
\affiliation{Institute for
Theoretical Physics (ITP) \& Spinoza Institute,
             Utrecht University, Leuvenlaan 4, Postbus 80.195,
              3508 TD Utrecht, The Netherlands}

\begin{abstract}
We consider the nonminimally coupled $\lambda\varphi^4$
scalar field theory in de Sitter space
and construct the renormalization group improved renormalized effective theory
at the one-loop level.
Based on the corresponding quantum Friedmann equation
and the scalar field equation of motion,
we calculate the quantum radiative
corrections to the scalar spectral index $n_s$, gravitational wave
spectral index $n_g$ and the ratio $r$ of tensor to scalar
perturbations. When compared with the standard (tree-level) values,
we find that the quantum contributions are suppressed by $\lambda N^2$
where $N$ denotes the number of $e$-foldings.
Hence there is an $N^2$ enhancement 
with respect to the na\"ive expectation, which is
due to the infrared enhancement of scalar vacuum fluctuations
characterising de Sitter space.
 Since observations constrain $\lambda$ to be very
small $\lambda \sim 10^{-12}$ and $N\sim 50-60$,
the quantum corrections in this inflationary model are unobservably small.

\end{abstract}

\pacs{98.80.-k, 98.80.Cq}



\maketitle

\section{Introduction}

Given the fact that we live in the era of precision cosmology, it
is important to establish a framework within which the quantum
radiative corrections for observables
 induced by the vacuum fluctuations of matter fields can be calculated.
Such radiative corrections may be important in some inflationary models.
In this paper we consider the nonminimal
${\lambda\varphi^4}$ inflationary model which includes a nonvanishing
coupling $\xi$ to the curvature scalar.
 We consider this model for simplicity; once the framework for
calculating the quantum (radiative) corrections is established, it
can be quite easily applied to other inflationary
models~\cite{ProkopecBilandzic}.

The minimally coupled ${\lambda\varphi^4}$ inflationary model
(with $\xi=0$) is already more than two standard deviations
disfavored by cosmological observations~\cite{Spergel:2006}.
However, for a certain choice of the coupling $\xi$ to the
background, the resulting nonminimally coupled scalar field model
can still match the experimental data~\cite{HwangNoh:1998,
KomatsuFutamase:1999, FakirUnruh:1990,Tsujikawa:2004my}, 
basically because
the model then produces the spectral index of
the massive chaotic inflaton model.
In this work we show that this is indeed the case,
but only for a rather limited values of $\xi$, namely for $\xi$ which satisfies
\begin{equation}
\frac{1}{8\tilde{N}}\ll |\xi | \ll\frac{1}{24}\,,\qquad (\xi<0) \,,
\label{condition on neg xi}
\end{equation}
where $\tilde{N}\simeq N+1$ and $N$ is the number of $e$-foldings.

If the condition $|dH/dt| \ll H^2$ is not
fulfilled, then our framework is not applicable, because we have
constructed the de Sitter invariant scalar field propagator with the
assumption that the Ricci scalar 
$R = 6(2H^2+d{H}/dt) \simeq 12H^2$, where $H=H(t)$ denotes the 
Hubble parameter. This condition is
fulfilled in most of inflationary models and thus does not present
a significant constraint to our model.

During inflation the amplitude of field correlators at the classical
level is suppressed by powers of the Hubble parameter, but the quantum
corrections to field correlators can depend on the whole history
of inflation, leaving hence the possibility that in a cumulative
manner quantum corrections can become important and even
detectable by future
experiments~\cite{BoyanovskydeVegaSanchez:2006,BoyanovskydeVegaSanchez:2005,Boyanovsky:2004ph,Weinberg:2005,Weinberg:2006}.
Such cumulative effects are claimed to be present in the analysis done
recently in Refs.~\cite{Sloth:2006az,Sloth:2006nu}. In our
analysis no such cumulative effects are present.

The quantum radiative corrections to slow-roll parameters in inflation have
been firstly calculated in
Refs.~\cite{BoyanovskydeVegaSanchez:2006,BoyanovskydeVegaSanchez:2005,Boyanovsky:2004ph}.
The authors begin by considering single field inflationary models
and subsequently generalize their analysis to include the inflaton
coupling to a light scalar and light fermionic field. While
Refs.~\cite{BoyanovskydeVegaSanchez:2006,BoyanovskydeVegaSanchez:2005}
consider quantum corrections to the equation of motion in momentum space, 
in this work we make use of the effective action techniques.
 Our results are in a qualitative disagreement with those of
Refs.~\cite{BoyanovskydeVegaSanchez:2006,BoyanovskydeVegaSanchez:2005,Boyanovsky:2004ph}.
One important difference is in that in their analysis the authors 
of~\cite{BoyanovskydeVegaSanchez:2006,BoyanovskydeVegaSanchez:2005,Boyanovsky:2004ph}
neglect the inflaton coupling to the background curvature (Ricci scalar), 
which within our framework yields the dominant contribution 
to the quantum radiative corrections during slow-roll inflation.
 A second important difference is that we made our analysis by using 
the de Sitter invariant propagator, while the proper analysis should be
conducted by making use of a scalar propagator suitable for
quasi-de Sitter spaces. Our method is based on the effective action 
approach. We arrive at our one-loop effective action
by making use of the position space propagator at coincidence.
Within this method we are able to use the well established
machinery of dimensional regularization, renormalization and
renormalization group improvement of our resulting effective field theory.
The authors of Ref.~\cite{Boyanovsky:2004ph} advocate the use of the
dynamical renormalisation group method (DRG)~\cite{Boyanovsky:2004gq}.
In that novel method the secular terms, which induce
a logarithmic growth (with conformal time) of the mode functions,
are resummed to yield the renormalisation group improved mode functions.
These improved mode functions exhibit regulated late time infrared divergences,
rendering the mode functions infrared finite. 

 More specifically, within our framework we obtain
 a quantum infrared enhancement to slow-roll parameters which is,
when compared to the classical values,
proportional to the number of $e$-foldings squared $N^2$. 
This enhancement is due to the scalar field mass generated by the coupling
to the background curvature scalar.
The authors
of \cite{BoyanovskydeVegaSanchez:2006,BoyanovskydeVegaSanchez:2005,Boyanovsky:2004ph}
obtain a quadratic enhancement but for the $\lambda\varphi^4$ model
without including the inflaton coupling to the Ricci scalar.
On the other hand, when compared to the classical value, 
the quantum corrections generated by the $\lambda\varphi^4$ 
interaction term are enhanced in our framework only linearly by $N$.

 It is by now a well established fact that quantum effects can
have quite a dramatic impact during inflation. An example is
the breakdown of conformal invariance for the photons
of scalar electrodynamics,
which has as a consequence a photon mass generation during inflation and
a generation of cosmological scale magnetic fields~\cite{Prokopec:PhotonMass}.
Similarly the quantum radiative effects break conformal invariance of the
fermions of the Yukawa theory in de Sitter space~\cite{ProkopecWoodard:2003}.
As a result fermions acquire a mass during
inflation~\cite{ProkopecWoodard:2003,Garbrecht:2006jm}, having as a consequence
a production of fermions during inflation and possibly
inflationary baryogenesis. Finally, the canonical coupling of gravitons
to fermions enhances the production of fermions during
inflation~\cite{MiaoWoodard:2005+2006}.

 The main result of our work is the quantum correction to
the scalar spectral index~(\ref{nsC}--\ref{nsQ2}).
The leading order contribution reads
\begin{equation}
(n_s-1)_Q = \frac{\lambda\tilde{N}(\xi-1/6)}{18\pi^2}\frac{\kappa}{(1-\kappa)^2(1-\frac23\kappa)^2} +{\cal O}(\lambda \ln(\tilde N))
\,,\qquad
\big(\kappa = 8\tilde N\xi, \tilde N\approx N+1-\xi/2\big)
\label{ns-1:Q}
\,,
\end{equation}
which is to be compared with the classical contribution,
\begin{equation}
(n_s-1)_C = -\frac{3}{\tilde N}\,\frac{1-\frac23\kappa}{1-\kappa}
\label{ns-1:C}
\,.
\end{equation}
The leading contribution~(\ref{ns-1:Q}) originates solely from the resummation
of the mass insertions generated by the coupling to the background curvature.
Since $\lambda \sim 10^{-12}$ the quantum contribution~(\ref{ns-1:Q})
is indeed too small to be observable. Note that the
condition~(\ref{condition for xi}) implies $-\tilde N/3\ll \kappa<1$,
such that $\kappa$ can be large and negative. If this is the case
the classical spectral index~(\ref{ns-1:C}) becomes consistent with
the current CMB data~\cite{Spergel:2006}.
We futhermore calculate the quantum corrections to the spectrum of curvature
perturbation, to the tensor spectral index, and to the ratio
of the tensor-to-scalar spectrum.

The present work is organized as follows. In
Section~\ref{Propagator in de Sitter space} we first recall the
basics of de Sitter space and then sketch the derivation of the de
Sitter invariant Chernikov-Tagirov propagator. In Section
\ref{Effective potential} we use this propagator and the
techniques of dimensional regularization and renormalization to
derive the one-loop improved effective potential for our theory.
This procedure requires one counterterm for the quartic
self-coupling constant $\lambda$ and one for the coupling to the
background $\xi$, which are defined at an arbitrary scale
$\varphi_0$. In Section~\ref{Renormalization group analysis} we
use the standard renormalization group (RG) techniques to improve
our effective potential. 
Having obtained the RG improved effective
potential, in Section \ref{Slow-roll parameters} we calculate the
corresponding quantum scalar field stress-energy tensor in the slow-roll
approximation. By making use of the quantum Friedmann equation and
of the scalar field equation of motion, we then develop our
framework within which the quantum radiative corrections from the
vacuum matter fluctuations to slow-roll parameters can be
calculated. In particular, we organize the quantum radiative
corrections to slow-roll parameters $\epsilon$ and $\eta$ into two
distinct parts. The first part arises from the one-loop
resummation of the mass insertions generated by the quartic
self-coupling in the presence of a scalar (inflaton) condensate,
while the second part arises from the resummation of the scalar mass
insertions generated by the coupling to the background. Both of
these quantum corrections are suppressed by the coupling constant
$\lambda$ but they are enhanced by the number of $e$-foldings squared.
Based on these results we then calculate the quantum radiative
corrections to the observables: the spectrum of curvature
perturbation and its spectral index, the tensor spectral index and
the ratio of tensor-to-scalar spectra. Finally, in
Section~\ref{Discussion} we summarize our results and discuss
their physical implications.

\section{Propagator in de Sitter space}
\label{Propagator in de Sitter space}

\subsection{de Sitter space}
\label{de Sitter space}

\begin{figure}[tbp]
\vskip -0.1in
\begin{center}
\epsfig{file=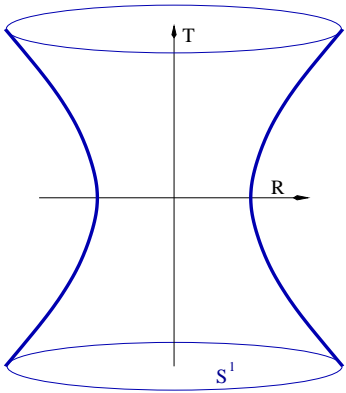,width=1.8in}
\end{center}
\lbfig{figure 1}
\vskip -0.3in
\caption[fig0]{%
\small
The embedding of de Sitter space into a five dimensional flat space-time.
The vertical line corresponds to the time coordinate, $X_0=T$,
and the radial coordinate $R=\sqrt{X_1^2+X_2^2+X_3^2+X_4^2}$.
 At each point $(T,R)$ there is a unit 3-sphere $S^3$, which is for
the sake of clarity represented by a circle $S^1$ erected at each
point $(T,R)$. The Hubble radius $R_H=1/H$
is the coordinate distance $R$ of the hyperboloid from the origin at $T=0$.
}
\end{figure}
A four dimensional de  Sitter space is perhaps best viewed as a 4-dimensional
hyperboloid embedded into the 5-dimensional Minkowski space-time
with the line element,
\begin{equation}
 ds_5^2 =  - dX_0 ^2 + dX_1^2 + dX_2^2 + dX_3^2 + dX_4^2
\label{dS:5 dim}
\,.
\end{equation}
The embedded hyperboloid of de Sitter space is shown in
Figure~\ref{figure 1}, and it is determined by
\begin{equation}
\label{dS:hyperboloid}
 - X_0^2 + X_1^2 + X_2^2 + X_3^2 + X_4^2  = \frac{1}{H^2}
\,,
\end{equation}
where $H$ denotes the Hubble parameter.
The symmetry group of de Sitter space,
$SO(1,4)$, is manifest by this embedding. One defines the de Sitter
invariant distance functions as,
\begin{equation}
 Z(X;X^\prime) = H^2 \sum_{A,B=0}^4\eta_{AB}X_A X^\prime_B
               = 1 - \frac{1}{2}Y(X;X^\prime)
\,,\qquad
\eta_{AB} = {\rm diag}(-1,1,1,1,1)
\,.
\label{invariant distance}
\end{equation}
We shall use the following flat 4-dimensional coordinates (which
cover 1/2 of the de Sitter manifold),
\begin{eqnarray}
  X_0 &=& \frac{1}{H}\sinh(Ht)+\frac{H}{2}\,{\rm e}^{Ht}\|\vec x\|^2
  \,,\qquad
            (-\infty < t < \infty)\,,
\nonumber\\
 X_i &=& {\rm e}^{Ht}x_i
  \,,\qquad\qquad\qquad\qquad\qquad\quad\!\!
            (-\infty < x_i < \infty,\; i=1,2,3)\,,\qquad
\nonumber\\
 X_4 &=& \frac{1}{H}\cosh(Ht)-\frac{H}{2}\,{\rm e}^{Ht}\|\vec x\|^2
\,,
\label{dS:flat coordinates}
\end{eqnarray}
in which the metric tensor reduces to the form
\begin{equation}
ds^2 = -dt^2 + a^2d\vec{x}^{\,2}\,,
\end{equation}
with the scale factor $a={\rm e}^{Ht}$. When written in terms of
conformal time $\eta$, defined as $ad\eta = dt$, the metric tensor
acquires the conformal form,
\begin{equation}
 g_{\mu\nu} = a^2(\eta) \eta_{\mu\nu}
\,,\qquad
 a = -\frac{1}{H\eta} \quad (\eta<0)
\,,\qquad
\eta_{\mu\nu} = {\rm diag}(-1,1,1,1)
\,.
\label{dS:scale factor}
\end{equation}

The invariant distance functions $Z(X;X^\prime) \equiv
z(x;x^\prime)$ and $Y(X;X^\prime) \equiv y(x;x^\prime)$ reduce in
these coordinates to the simple form,
\begin{equation}
 z(x;x^\prime) = 1 - \frac 12 y(x;x^\prime)
\,,\qquad
 y(x;x^\prime) = aa^\prime H^2 \Delta x^2\,,
\label{invariant distance:2}
\end{equation}
with $a = a(\eta) = -1/(H\eta)$,
$a^\prime = a(\eta^\prime) = -1/(H\eta^\prime)$, and
\begin{equation}
\Delta x^2(x;x^\prime) = -(|\eta-\eta^\prime|-i\varepsilon)^2
                + \|\vec x-\vec x^{\,\prime}\|^2
\,,
\label{Delta x}
\end{equation}
where (for a later use)  we introduced the infinitesimal parameter
$\varepsilon>0$, which defines how the poles of the propagator
(discussed in the next section) contribute.
In these coordinates the curvature of spatial sections vanishes,
and thus they are also known as flat (Euclidean) coordinates, in which
de Sitter space appears as uniformly expanding.

 By solving the relevant geodesic equations for $x^0$ and $x^i$ ($i=1,2,3$)
one can show that the de Sitter invariant distance function
$y=y(x;x^\prime\,)$ is related to
the geodesic distance $\ell=\ell(x;x^\prime\,)$ by the following simple
relation,
\begin{equation}
 y(x;x^\prime\,) = 4 \sin^2\left(\frac12 H\ell(x;x^\prime\,)\right)
\,.
\label{geodesic distance}
\end{equation}

\subsection{Scalar propagator in de Sitter space}
\label{Scalar propagator in de Sitter space}

The dynamics of the scalar field are specified by the following
tree-level action,
\begin{eqnarray}
 S_\varphi = \int d^4 x \sqrt{-g} {\cal L}_\varphi
\,,
\label{scalar tree action}
\end{eqnarray}
with the Lagrangean,
\begin{eqnarray}
\sqrt{-g} {\cal L}_\varphi =
 \sqrt{-g}
   \Big(
   - \frac12 g^{\mu\nu}(\partial_\mu\varphi)(\partial_\nu\varphi)
   - V_b(\varphi)
 \Big)
\,,
\label{scalar lagrangean}
\end{eqnarray}
where
\begin{equation}
 V_b(\varphi) = \frac12 \xi_b {\cal
 R}\varphi^2 + \frac{\lambda_b}{4!} \varphi^4 \,.
\label{scalar:potential}
\end{equation}
In the above expression $\lambda_b$ and $\xi_b$ are the bare values of
the quartic self-coupling and the nonminimal coupling
to the Ricci curvature scalar ${\cal R}$, respectively, and
$g={\rm det}(g_{\mu\nu})$. For simplicity we set the bare scalar mass
$m_b=0$. The theory~(\ref{scalar tree action}--\ref{scalar:potential})
is a simplified version of the Yukawa theory in de Sitter background
studied in Refs.~\cite{Prokopec:2006,MiaoWoodard:2006}.
An early related work can be found in Ref.~\cite{CandelasRaine:1975}.

The scalar propagator in a curved background space
can be defined as the expectation value,
\begin{equation}
 i\Delta(x;x^\prime) = \langle x|
                  \frac{i}{\sqrt{-g}(\Box - m_\varphi^2 - \xi_b{\cal R}_D)}
                  |x^\prime\rangle
\,,\qquad \Big(m_\varphi^2(\varphi)
               \equiv \frac{\lambda_b}{2}\varphi^2\Big)
\,,
\label{de Sitter scalar propagator}
\end{equation}
where $|x\rangle$ is the eigenstate of the position operator $\hat x$
({\it i.e.} $\hat x |x\rangle = x|x\rangle$),
 $\Box=(-g)^{-1/2}\partial_\mu (-g)^{1/2}g^{\mu\nu}\partial_\nu$ denotes
the d'Alambertian and $g={\rm det}[g_{\mu\nu}]$.
This Feynman propagator solves the following equation in de Sitter space
in general $D$ space-time dimensions
(needed for dimensional regularization and renormalization),
\begin{equation}
\sqrt{-g} (\Box - m_\varphi^2 - \xi_b{\cal
R}_D)i\Delta(x;x^\prime)
          = i \delta^D(x-x^\prime)
\,,
\label{Dirac propagator equation}
\end{equation}
where $\delta^D(x-x^\prime)$
is the $D$-dimensional Dirac $\delta$-distribution,
${\cal R}_D = D(D-1)H^2$ is the Ricci scalar in a $D$ dimensional
de Sitter space, and $H$ is the Hubble parameter.

 The de Sitter invariant form of (\ref{Dirac propagator equation})
is~\cite{ChernikovTagirov:1968,Tagirov:1972,BunchDavies:1978,ProkopecPuchwein:2003}
\begin{equation}
     \bigg[
         (1-z^2)\frac{d^2}{dz^2}
       - Dz\frac{d}{dz}
       - \frac{m_\varphi^2+\xi_b R_D}{H^2}
    \bigg]i G(y)
       = \frac{i \delta^D(x-x^\prime)}{H^2 a^{D}}
\,,
\label{dS:scalar propagator}
\end{equation}
where the invariant propagator is defined as, $iG(y) =
i\Delta(x;x^\prime)$. Here we made use of
Eqs.~(\ref{invariant distance:2}--\ref{Delta x}) and of
\begin{equation}
 \partial_\mu \equiv (\partial_\mu z)\frac{d}{dz}
                = - \frac12 Ha\Big(
                                  \delta_\mu^{\;0}y + 2 a^\prime H \Delta x_\mu
                               \Big) \frac{d}{dz}
\,.
\label{partial-mu:z}
\end{equation}

The properly normalized de Sitter invariant solution of
Eq.~(\ref{dS:scalar propagator}),
which near the light-cone and in the massless limit reduces to the Hadamard
form,
\begin{equation}
  iG(y)\; \stackrel{y\rightarrow 0}{\longrightarrow}\;
 \frac{H^{D-2}}{(2\pi)^{D/2}}\Gamma\Big(\frac{D}{2}-1\Big)
               \frac{1}{y^{\frac{D}{2}-1}}
   + {\cal O}\Big(y^{2-D/2},y^0\Big)
\,,
\label{dS:Hadamard form}
\end{equation}
is unique,
\begin{equation}
    iG(y) = \frac{H^{D-2}}{(4\pi)^{\frac{D}{2}}}
           \frac{\Gamma\Big(\frac{D-1}{2}+\nu_D\Big)
                     \Gamma\Big(\frac{D-1}{2}-\nu_D\Big)}
                {\Gamma\Big(\frac{D}{2}\Big)}
                 \phantom{\;}_2F_1\Big(
                                   \frac{D-1}{2}+\nu_D,
                                   \frac{D-1}{2}-\nu_D;
                                   \frac{D}{2};
                                   1-\frac{y}{4}
                              \Big)
\,,
\label{dS:scalar propagator:2}
\end{equation}
where
\begin{equation}
    \nu_D = \bigg({\frac{(D-1)^2}{4}
               -  \frac{m_\varphi^2 + \xi_b R_D}{H^2}}
            \bigg)^\frac 12
\,.
\label{dS:scalar propagator:nu}
\end{equation}
This is the Chernikov-Tagirov propagator for de Sitter
space~\cite{ChernikovTagirov:1968,Tagirov:1972}
generalized to $D$ space-time dimensions.
 The pole prescription defined by the $i\varepsilon$-prescription in
Eq.~(\ref{Delta x})
implies that the propagator~(\ref{dS:scalar propagator:2})
corresponds to the time-ordered (Feynman) propagator.
For a discussion of other propagators relevant for expanding space-times
in the Schwinger-Keldysh {\it in-in} formalism we refer
to~\cite{ProkopecPuchwein:2003,Weinberg:2005}.

\bigskip

\section{Effective potential}
\label{Effective potential}

\begin{figure}[tbp]
\centerline{\hspace{.in}
\epsfig{file=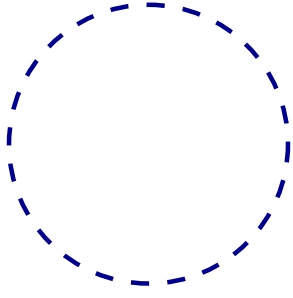, width=1.5in}
}
\lbfig{figure 2}
\vskip -0.1in
\caption{\small
The one-loop diagram (vacuum bubble) contributing to the scalar effective
theory~(\ref{effective action:1 loop}) in a curved background.
}
\end{figure}
The one-loop effective action for a real scalar field reads,
\begin{equation}
 \Gamma_\varphi[g_{\mu\nu},\varphi] \equiv \int d^Dx \sqrt{-g}{\cal L}_\varphi
  = S_{\rm HE}[g_{\mu\nu}] + S_\varphi[g_{\mu\nu},\varphi]
   + \frac i2{\rm Tr}\ln\Big(
                          \sqrt{-g}\big(\Box - m_{\varphi}^2-\xi_b{\cal R}_D\big)
                        \Big)
\,,
\label{effective action:1 loop}
\end{equation}
where ${\rm Tr}$ refers to the space-time integration $\int d^Dx$,
$S_\varphi$ is the tree-level scalar field
action~(\ref{scalar tree action}) and $S_{\rm HE}$ denotes
the Hilbert-Einstein action,
\begin{equation}
 S_{\rm HE} = -\frac{1}{16\pi G_N} \int d^Dx \sqrt{-g}{\cal R_D}
\,.
\label{Hilbert-Einstein}
\end{equation}
The last term in Eq.~(\ref{effective action:1 loop}) represents
the one-loop contribution to the effective action $\delta_1 {\cal
L}_\varphi$, whose graphical representation is shown in
Figure~\ref{figure 2}.
 We shall now evaluate the general expression~(\ref{effective action:1 loop})
in de Sitter space. It is convenient to differentiate the
one-loop contribution $\delta_1 {\cal L}_\varphi$
with respect to the scalar mass,
\begin{eqnarray}
  \frac{\partial \delta_1{\cal L}_\varphi}{\partial m_\varphi^2} &=&
        \frac i2 \bigg\langle x\bigg|
               \frac{-1}{\sqrt{-g}\big(\Box - m_{\varphi}^2-\xi_b{\cal R}_D\big)}
            \bigg| x\bigg\rangle
        = -\frac12\, i{\Delta}(x;x)
\label{effective action:3} \,.
\end{eqnarray}
Now making use of Eq.~(\ref{dS:scalar propagator:2}) one obtains,
\begin{eqnarray}
  \frac{\partial \delta_1{\cal L}_\varphi}{\partial m_\varphi^2}
   &=& -\frac12 iG(y)|_{y\rightarrow 0}
      = -\frac12\frac{H^{D-2}}{(4\pi)^{D/2}}\Gamma\Big(1-\frac{D}{2}\Big)
           \frac{\Gamma\Big(\frac{D-1}{2}+\nu_D\Big)
                  \Gamma\Big(\frac{D-1}{2}-\nu_D\Big)}
                {\Gamma\Big(\frac{1}{2}+\nu_D\Big)
                    \Gamma\Big(\frac{1}{2}-\nu_D\Big)}
\,.
\label{effective action:scalars}
\end{eqnarray}
Separating the divergent and finite contributions
in~(\ref{effective action:scalars}) yields
\begin{eqnarray}
  \frac{\partial \delta_1{\cal L}_\varphi}{\partial m_\varphi^2}
   &=& -\frac12\frac{H^{D-2}}{(4\pi)^{D/2}}\Gamma\Big(1\!-\!\frac{D}{2}\Big)
           \Big(\frac{m_\varphi^2}{H^2}
              - (D\!-\!2) + \xi_b D(D\!-\!1)
           \Big)
\nonumber\\
&&  -\,
\frac{H^2}{32\pi^2}\Big(\frac{m_\varphi^2}{H^2}-2(1-6\xi_b)\Big)
          \Big[\psi\Big(\frac{1}{2}\!+\!\nu\Big)
              +\psi\Big(\frac{1}{2}\!-\!\nu\Big)\Big]
\,,\qquad
\label{effective action:scalars:2}
\end{eqnarray}
where we made use of,
\begin{equation}
\Gamma\Big(1-\frac{D}{2}\Big) = \frac{2}{D-4} - (1-\gamma_E) + {\cal O}(D-4)
\,,
\label{Gamma 1-D/2}
\end{equation}
$\gamma_E \simeq 0.577$ is the Euler constant,
${\cal R}_{D=4}\equiv {\cal R} =12H^2$, and
\begin{equation}
\nu = \Big(\frac14 - \frac{m_\varphi^2}{H^2} +
2(1\!-\!6\xi_b)\Big)^{\frac12} \,,
\label{def:nu}
\end{equation}
where $m_\varphi ^2 = \lambda \varphi^2/2$ is defined in 
Eq.~(\ref{de Sitter scalar propagator}).
When integrated, Eq.~(\ref{effective action:scalars:2}) gives
the following contribution to the effective Lagrangean,
\begin{eqnarray}
\delta_1 {\cal L}_\varphi
    &=& -\frac{1}{2(4\pi)^{D/2}}\Gamma\Big(1\!-\!\frac{D}{2}\Big)
           \Big(\frac12H^{D-4}m_\varphi^4
          - \Big[(D\!-\!2)-\xi_b D(D\!-\!1)\Big]H^{D-2}m_\varphi^2\Big)
\label{effective action:scalars:3}
\\
&&\hskip 0.cm
     - \frac{H^{2}}{32\pi^2}
        \!\int\! dww
              \Big[\psi\Big(\frac{1}{2}\!+\!\nu\Big)
                              + \psi\Big(\frac{1}{2}\!-\!\nu\Big)\Big]\,,\qquad
w = \frac{m_\varphi^2}{H^2} +12\left(\xi_b-\frac16\right)\,,
\nonumber
\end{eqnarray}
where the integral is an indefinite integral.

In order to renormalize our Lagrangean ${\cal L}_\varphi $ we will
add to it the counterterms $\lambda_0$ and $\xi_0$ and apply
the renormalization conditions which will determine the finite parts
of those counterterms,
\begin{eqnarray}
{\cal L}_{\varphi,\,{\rm ren}} & = & - \frac12
g^{\mu\nu}(\partial_\mu\varphi)(\partial_\nu\varphi) -
V_{\mathrm{ren}}(\varphi)
\,,
\label{eff:lagrangean}
\end{eqnarray}
where now
\begin{eqnarray}
V_{\mathrm{ren}}(\varphi) & = & \frac{\lambda_b}{4!}\varphi^{4} +
\frac12 \xi_b \mathcal{R}\varphi^{2} +
\frac{\lambda_0}{4!}\varphi^{4} + \frac12 \xi_0
\mathcal{R}\varphi^{2} - \delta_1 {\cal L}_\varphi,
\label{Vren}
\end{eqnarray}
and $\delta_1 {\cal L}_\varphi$ is given by
Eq.~(\ref{effective action:scalars:3}).

We renormalize our Lagrangean at an arbitrary scale $\varphi_0$,
\begin{eqnarray}
\left.
\frac{\delta^4V}{\delta\varphi^4}\right|_{\varphi=\varphi_0} &=&
\lambda_b = \lambda_b + \lambda({\varphi_0},H^2) -
\left. \frac{\delta^4(\delta_1{\cal L})}{\delta\varphi^4}\right|_{\varphi=\varphi_0}\,,\nonumber\\
\left.
\frac{\delta^3V}{\delta(H^2)\delta\varphi^2}\right|_{\varphi=\varphi_0}
&=& D(D\!-\!1)\xi_b = D(\!D-\!1)\xi_b +
D(\!D-\!1)\xi(\varphi_0,H^2) - \left. \frac{\delta^3(\delta_1{\cal
L})}{\delta(H^2)\delta\varphi^2}\right|_{\varphi=\varphi_0}\,,
\label{renormalization conditions}
\end{eqnarray}
which yields
\begin{eqnarray}
\lambda({\varphi_0},H^2) &=&
-\frac{3}{2(4\pi)^{D/2}}\Gamma(1-\frac{D}{2})\mu^{D-4}\lambda_b^2
\nonumber
\\
& &
-\frac{3\lambda_b^2}{32\pi^2}\left[\ln\left(\frac{\lambda_b\varphi_0^2+24H^2(\xi_b-1/6)}{{2\mu^2}}\right)
+\frac73 +
\frac43\frac{\lambda_b\varphi_0^2}{\lambda_b\varphi_0^2+24H^2(\xi_b-1/6)}
\right]
\,,
\nonumber\\
\xi(\varphi_0,H^2)
  &=&\frac{\lambda_b}{2(4\pi)^{D/2}}
           \Gamma(1-\frac{D}{2})\left(\frac{D-2}{D(D-1)}-\xi_b\right)\mu^{D-4}
\nonumber\\
    & &-\frac{\lambda_b}{32\pi^2}\left[\left(\xi_b-\frac16\right)\ln\left(
    \frac{\lambda_b\varphi_0^2+24H^2(\xi_b-1/6)}{{2\mu^2}}\right)
         +3\left(\xi_b-\frac16\right)-\frac{1}{36}\right]
\,.
\end{eqnarray}
From (\ref{renormalization conditions}) it follows that the
counterterms $\lambda_0$ and $\xi_0$ are given by
\begin{eqnarray}
\lambda_0 \equiv \lambda({\varphi_0},0) &=&
-\frac{3}{2(4\pi)^{D/2}}\Gamma\left(1-\frac{D}{2}\right)\mu^{D-4}\lambda_b^2
-\frac{3\lambda_b^2}{32\pi^2}\left[\ln\left(\frac{\lambda_b\varphi_0^2}{2\mu^2}\right)+\frac{11}{3}\right]\nonumber\\
\xi_0 \equiv \xi({\varphi_0},0) &=&
\frac{\lambda_b}{2(4\pi)^{D/2}}
                \Gamma\Big(1-\frac{D}{2}\Big)
                       \bigg[\frac{D\!-\!2}{D(D-1)}-\xi_b\bigg]
                       \mu^{D-4}
\nonumber\\
         & & -\frac{\lambda_b}{32\pi^{2}}\bigg[\Big(\xi_b-\frac{1}{6}\Big)
                       \ln\Big(\frac{\lambda_b\varphi_0^{2}}{2\mu^{2}}\Big)
                       +3\Big(\xi_b-\frac{1}{6}\Big)-\frac{1}{36}\bigg]
\,.
\label{counterterms}
\end{eqnarray}
Given that $H$ and $\varphi$ are dynamical fields, the counterterm parameters
$\lambda_0$ and $\xi_0$ must be independent of $H$ and $\varphi$,
which is indeed satisfied by~(\ref{counterterms}).
 Now making use of Eqs.~(\ref{effective action:scalars:3}),
(\ref{Vren}) and~(\ref{counterterms}) we can calculate the renormalized
Lagrangean~(\ref{eff:lagrangean}). The result is,
\begin{eqnarray}
 {\cal L}_{\varphi,\, \mathrm{ren}}
  &=& -\frac12g^{\mu\nu}\partial_\mu\varphi\partial_\nu\varphi
      - \frac{\varphi^4}{4!}\bigg\{\lambda_b+\frac{3\lambda_b^{2}}{32\pi^{2}}
      \bigg[\ln\Big(\frac{2H^{2}}{\lambda_b\varphi_0^{2}}\Big)-\frac{11}{3}\bigg]\bigg\}
\nonumber\\
  & &    - \frac12{\cal R}\varphi^2\bigg\{\xi_b+\frac{\lambda_b}{32\pi^{2}}\bigg[\Big(\xi_b-\frac{1}{6}\Big)
                       \ln\Big(\frac{2H^{2}}{\lambda_b\varphi_0^{2}}\Big)
                       -3\Big(\xi_b-\frac{1}{6}\Big)+\frac{1}{36}\bigg]\bigg\}
\nonumber
\\
&&\hskip 0.cm
     - \frac{H^4}{32\pi^2}
        \!\int\! dww
              \Big[\psi\Big(\frac{1}{2}\!+\!\nu\Big)
                              + \psi\Big(\frac{1}{2}\!-\!\nu\Big)\Big]
\,.
\label{effective action:scalars:4}
\end{eqnarray}
This is the fully renormalized effective Lagrangean. We now
consider the two asymptotic forms of (\ref{effective
action:scalars:4}), first the ultraviolet (UV) limit.

The following asymptotic expansion of the di-gamma function,
$\psi(z) = (d/dz)[\ln(\Gamma(z))]$, is then useful ({\it cf.}
Eq.~(8.344) in \cite{GradshteynRyzhik:1965}),
\begin{equation}
  \psi(z) = \ln(z) - \frac{1}{2z} - \frac{1}{12z^2}
            + \frac{1}{120z^4} + {\cal O}(z^{-6})
\,,
\label{psi:asymptotic:0}
\end{equation}
such that
\begin{eqnarray}
 \psi\Big(\frac 12 + \nu\Big) + \psi\Big(\frac 12 - \nu\Big)
  &=& \ln(w) - \frac{1}{3w} - \frac{1}{15w^2} + {\cal O}(1/w^{3})
\,,\qquad \nu^2 = \frac14 - w
\,.
\label{psi:asymptotic expansion}
\end{eqnarray}
Upon evaluating the integral in~(\ref{effective action:scalars:2})
one obtains
\begin{equation}
 -\frac{H^4}{32\pi^2}
   \bigg\{
          \frac{w^2}{2}\Big[\ln(w)-\frac12\Big]
       -  \frac{w}{3}
       -  \frac{1}{15}\ln(w) + {\cal O}(1/w)
   \bigg\}
\,.
\end{equation}
Taking account of this, we can recast Eq.~(\ref{effective action:scalars:4})
to the form,
\begin{eqnarray}
 {\cal L}_\varphi
  &=& -\frac12g^{\mu\nu}\partial_\mu\varphi\partial_\nu\varphi
      - \frac{\varphi^4}{4!}\left\{\lambda_b+\frac{3\lambda_b^{2}}{32\pi^{2}}
      \left[\ln\left(\frac{\lambda_b\varphi^{2}+24H^{2}
      \left(\xi_b-\frac16\right)}{\lambda_b\varphi_0^{2}}\right)-\frac{25}{6}\right]\right\}
\nonumber\\
  & & -\frac12{\cal R}\varphi^2\left\{\xi_b+\frac{\lambda_b}{32\pi^{2}}\left[\Big(\xi_b-\frac16\Big)
                       \ln\left(\frac{\lambda_b\varphi^{2}+24H^{2}\left(\xi_b-\frac16\right)}{\lambda_b\varphi_0^{2}}\right)
                       -\frac72\left(\xi_b-\frac16\right)+\frac{1}{72}\right]\right\}
\label{effective action:scalars:5}
\nonumber\\
& &\hskip 0.cm
    -\, \frac{{\cal R}^2}{64\pi^2}
             \left\{
               \left(\xi_b-\frac16\right)^2
                   \left[\ln\left(\frac{\lambda_b\varphi^2}{2H^{2}}
                         +12\left(\xi_b-\frac16\right)\right)
                         -\frac12
                   \right]
             -\frac{1}{18}\left(\xi_b-\frac16\right)\right.
\nonumber\\
& &
             -\, \left.\frac{1}{1080}\ln\left(\frac{\lambda_b\varphi^2}{2H^{2}}
             +12\left(\xi_b-\frac16\right)\right)\right\}+{\cal O}({\cal R}^3).
\end{eqnarray}
This is the UV form of (\ref{effective action:scalars:4}). In the
limit $H^2\rightarrow 0$ to flat Minkowski spacetime, the first
line in (\ref{effective action:scalars:5}) reproduces the
classical Coleman-Weinberg result~\cite{ColemanWeinberg:1973}.

 To complete the analysis of the effective
Lagrangean~(\ref{effective action:scalars:4}) we still need to
consider the small field limit of the integral in~(\ref{effective
action:scalars:4}). Let us first consider the expansion which is
applicable around the poles of the di-gamma function, which are
located at
\begin{equation}
  \nu_n = \frac12 + n\,,\qquad n=0,1,2,3,\dots
\,.
\label{poles of psi}
\end{equation}
This implies that the poles are located at
\begin{equation}
  w_n = -n(n+1)
\label{w_n}
\end{equation}
and in the vicinity of the poles we can write
\begin{equation}
  w = w_n +\delta w
\,.
\label{w}
\end{equation}
With the above definitions, a conformally coupled scalar field with $\xi= 1/6$
corresponds to $n=0$ and a minimally coupled scalar field
with $\xi = 0$ corresponds to $n = 1$. More generally we have,
\begin{equation}
\xi_n = \frac{(1-n)(2+n)}{12}
\,, \qquad (n\geq 0)
\,,
\label{xi_n}
\end{equation}
such that for $n>1$ all $\xi_n<0$.
In particular, $\xi_2 = -1/3$, $\xi_3 = -5/6$, {\it etc.}
Now we can expand $\nu$ as
\begin{equation}
\nu = \nu_n - \frac{\delta w}{2\nu_n}-\frac18\frac{(\delta
w)^2}{\nu_n^{3}}
\end{equation}
and by making use of
\begin{eqnarray}
 \psi\Big(\frac12 -\nu\Big) =  \psi\Big(n + \frac32 - \nu\Big)
                             - \sum_{\ell=0}^{n} \frac{1}{\frac12 + \ell - \nu}
\label{psi-log}
\end{eqnarray}
we can finally write
\begin{eqnarray}
 w\left[\psi\Big(\frac12 + \nu\Big) + \psi\Big(\frac12
 -\nu\Big)\right]
         = \frac{a_n}{\delta w} + b_n + c_n\delta w
   + {\cal O}((\delta w)^2)
\,.
\label{psi-log:2}
\end{eqnarray}
In the above expressions
\begin{eqnarray}
a_n & = & n(n+1)(2n+1)
\,,
\nonumber\\
b_n & = & -2n(n+1)\psi(n+1)-(2n+1)-\frac{n(n+1)}{2n+1}
\,,
\nonumber\\
c_n & = & \frac{1}{2n+1}+2\psi(n+1)
     +\frac{2n(n+1)}{2n+1}\left(-\psi'(1)+\psi'(n+1)-\frac{1}{2(2n+1)^2}\right)
\,,
\label{abc}
\end{eqnarray}
and we made use of
\begin{eqnarray}
 \sum_{\ell=1}^n\frac{1}{\ell} = \psi(n+1)-\psi(1)
\,,\quad
 \sum_{\ell=1}^n\frac{1}{\ell^2} = -\psi^\prime(n+1)+\psi^\prime(1)
\,,
\label{psi-log:3}
\end{eqnarray}
where $\psi(1)=-\gamma_E = 0.577215...$ is the Euler constant,
$\psi^\prime(1) = \pi^2/6$ and $\psi^\prime(z+1) = \psi^{\,\prime}(z) - 1/z^2$.

 We can now write the infrared limit of the
 renormalized Lagrangean~(\ref{effective action:scalars:4}).
From Eqs.~(\ref{psi-log:2}--\ref{abc}) it follows that
the integral in the last line of Eq.~(\ref{effective action:scalars:4})
has the infrared limit,
\begin{equation}
 -\frac{H^4}{32\pi^2}\left[
                          a_n\ln(\delta w_n)
                        + b_n(\delta w_n)
                        + \frac12c_n(\delta w_n)^2
                        + {\cal O}\Big((\delta w_n)^3\Big)
                    \right]
\,,
\label{Veff:IR integral}
\end{equation}
where
\begin{equation}
  \delta w_n = \frac{\lambda\varphi^2}{2H^2} + 12 \delta\xi_n
 \,,\qquad \delta\xi_n = \xi_b + \frac{1}{12}(n-1)(n+2)\ll 1
\,
\label{delta w_n}
\end{equation}
and $a_n,b_n$ and $c_n$ are defined in Eq.~(\ref{abc}).

\section{Renormalization group analysis}
\label{Renormalization group analysis}

In deriving expression (\ref{effective action:scalars:4}) we have
introduced an arbitrary renormalization scale $\varphi_0$ by definining
the renormalization conditions (\ref{renormalization
conditions}). From this definition an arbitrary scale
$\varphi_{0}$ enters into the expressions for the counterterms
(\ref{counterterms}) and hence into the renormalized Lagrangean
(\ref{effective action:scalars:4}). However, as it was stressed in
the classic paper of Coleman and Weinberg in 1973, the change of
the renormalization scale can only change the definitions of coupling
constants, not the physics~\cite{ColemanWeinberg:1973}.

By applying the same reasoning in our case, we arrive at the following
conclusion: a small change in $\varphi_0$ in the expression
for the physical quantity of interest can always be compensated for
by an appropriate small change in $\lambda$ and $\xi$. The
convenient way of expressing this statement is
\begin{equation}
\left(\varphi_0\frac{\partial}{\partial\varphi_0} +
\beta_\lambda\frac{\partial}{\partial\lambda} +
\beta_\xi\frac{\partial}{\partial\xi}\right)V_{\mathrm{eff}}(\varphi_0,
\lambda, \xi, \varphi) = 0, \label{callan-symanzik}
\end{equation}
which is the standard Callan-Symanzik (CS) equation written for the
theory at hand. The renormalization group
functions $\beta_\lambda$ and $\beta_\xi$ are given by the
following relations,
\begin{eqnarray}
\beta_\lambda &=&
-\varphi_0\left.\frac{\partial\lambda_0}{\partial\varphi_0}\right|_{\lambda_b}\,,\nonumber\\
\beta_\xi &=&
-\varphi_0\left.\frac{\partial\xi_0}{\partial\varphi_0}\right|_{\xi_b}\,.
\end{eqnarray}
Within the one-loop approximation the renormalization group
functions $\beta_\lambda$ and $\beta_\xi$ are uniquely determined
as the coefficients of the divergent logarithmic terms appearing in
the counterterms $\lambda_0$ and $\xi_0$~(\ref{counterterms}). It
follows (writing $\lambda_b$ and $\xi_b$ from now on as $\lambda$
and $\xi$, respectively):
\begin{eqnarray}
\beta_\lambda & = & \frac{3\lambda^{2}}{16\pi^2},\nonumber\\
\beta_\xi     & = &
\frac{\lambda}{16\pi^2}\left(\xi-\frac16\right).
\label{betafunctionsfinal}
\end{eqnarray}
These expressions for $\beta_\lambda$ and $\beta_\xi$ we will use
to determine the running of $\lambda$ and $\xi$ with the scale $\varphi_0$.

We shall now solve the Callan-Symanzik
equation~(\ref{callan-symanzik}). From the theory of partial
differential equations we can make use of the method of
characteristics~\cite{Buchbinder:1992rb,Elizalde:1993ew,cc,bkmn}. 
Applying this method to~(\ref{callan-symanzik}) 
we can write down the solution to the
Callan-Symanzik equation~(\ref{callan-symanzik}) as
\begin{equation}
V_{\mathrm{eff}}(\varphi_0, \lambda, \xi, \varphi) =
V_{\mathrm{eff}}(\varphi_0(t), \lambda (t), \xi(t), \varphi (t)),
\label{solutionofC-S}
\end{equation}
where $\varphi_0(t), \lambda(t), \xi(t), \varphi(t)$ are the running
parameters. The $t$-dependence of the running parameters is given
(to the order we are working in) by the following
differential equations:
\begin{eqnarray}
\frac{d\varphi_0(t)}{dt} & = & \varphi_0(t),
\hspace{2cm}
\frac{d\varphi(t)}{dt} = 0,
\nonumber\\
\frac{d\lambda(t)}{dt} & = & \beta_{\lambda}(\lambda (t)),
\hspace{1.52cm} \frac{d\xi(t)}{dt} =  \beta_{\xi}(\xi(t),
\lambda(t)). \label{differentialequations}
\end{eqnarray}
The boundary conditions (at $t=0$) are,
\begin{eqnarray}
\varphi_0(0) & = & \varphi_0, \hspace{1cm}
\varphi(0) = \varphi,\nonumber\\
\lambda(0) & = & \lambda, \hspace{1.27cm} \xi(0) = \xi.
\end{eqnarray}
The solutions of the first two differential equations
in~(\ref{differentialequations}) are trivial and read,
\begin{equation}
\varphi_0^2(t) = \varphi_0^2e^{2t}, \hspace{1cm} \varphi(t) = \varphi
\,.
\end{equation}
When combined with the previous results~(\ref{betafunctionsfinal}),
the last two differential equations
in~(\ref{differentialequations}) are solved by
\begin{equation}
\lambda(t) = \frac{\lambda}{1-\frac{3\lambda}{16\pi^2}t},
\label{running:lambda}
\end{equation}
and
\begin{equation}
\xi(t) = \frac16 + (\xi -
\frac16)\left(\frac{\lambda(t)}{\lambda}\right)^{1/3}
\,,
\label{running:xi}
\end{equation}
where here $\lambda=\lambda(0)$ and $\xi=\xi(0)$.
 These solutions imply that our model exhibits an infrared
fixed point, where the coupling constants are $\lambda_{\rm FP} = 0$
and $\xi_{\rm FP} = 1/6$. In this limit the theory possesses
an enhanced symmetry (conformal symmetry) and
it can be reduced to a pure metric theory
(see Ref.~\cite{Tsamis:1984hh} for a nonperturbative proof of this statement).

We stress that the parameter $t$ in the above relations is completely
arbitrary. The basic idea of the renormalization group (RG) improvement
of an effective potential is that we can choose $t$ in such a way that
the perturbation series for the effective potential converges more
rapidly. Indeed by suitably choosing $t$ one can
extend the range of validity of the effective theory to a larger
range of the dynamical quantities $H$ and $\varphi$
by replacing the perturbative expression on the left-hand side
of~(\ref{solutionofC-S}) by its right-hand side. This is intimately related to
the fact that the unimproved expression for the effective
potential is actually valid only for $\varphi$'s sufficiently close
to $\varphi_0$. Since the change of the arbitrary scale $\varphi_0$ corresponds
just to a reparametrization of the coupling constants
within our theory, the unimproved effective potential
is valid only near $\varphi \sim \varphi_0$ and $H\sim 0$
and thus not a very useful quantity. If we, on the other
hand, require that the effective theory does
not depend on $\varphi_0$, then the improved effective potential
(which solves the Callan-Symanzik equation)
remains valid whenever the coupling constants are small.

Since the perturbation series for the effective potential is
characterized by the occurrence of powers of logarithmic terms, we
choose\footnote{The choice (\ref{t}) is not unique. However, this
is the unique choice for which $\varphi_0\partial_{\varphi_0} =
-\partial_t$ when $H^2 \rightarrow 0$ in the CS equation
(\ref{callan-symanzik}). For any other choice of $t$ there is an
additional prefactor in front of $\partial_t$, and after dividing
the CS equation with that prefactor one can solve it \textit{as if}
the $\beta$ functions are modified, which in turn will modify the
functions $\lambda(t)$ and $\xi(t)$. This in principle leads to a
different effective RG improved theory, which however differs from
the one we use here only at higher orders in the coupling constants.}
\begin{equation}
t = \ln\left(\frac{\varphi}{\varphi_{0}}\right)\,.\label{t}
\end{equation}
The improved expression for the renormalized effective potential
now becomes
\begin{eqnarray}
V_{\mathrm{RG}}(\varphi)
&=&\frac{\varphi^4}{4!}\bigg\{\lambda+\frac{3\lambda^{2}}{32\pi^{2}}
      \bigg[\ln\Big(\frac{2H^{2}}{\lambda\varphi^{2}}\Big)-\frac{11}{3}\bigg]\bigg\}
\nonumber\\
  & &    + 6H^2\varphi^2\bigg\{\xi+\frac{\lambda}{32\pi^{2}}\bigg[\Big(\xi-\frac{1}{6}\Big)
                       \ln\Big(\frac{2H^{2}}{\lambda\varphi^{2}}\Big)
                       -3\Big(\xi-\frac{1}{6}\Big)+\frac{1}{36}\bigg]\bigg\}
\label{improved effective potential}
\\
&&\hskip 0.cm
     +\frac{H^4}{32\pi^2}
        \!\int\! dww
              \Big[\psi\Big(\frac{1}{2}\!+\!\nu\Big)
                              + \psi\Big(\frac{1}{2}\!-\!\nu\Big)\Big]
\nonumber
\,,
\end{eqnarray}
where $\lambda$ and $\xi$ are now $t$-dependent according to
(\ref{running:lambda}) and (\ref{running:xi}). We remark that even
after the improvement logarithmic terms still appear in the
expression for the effective potential. As it can be seen
from Eq.~(\ref{effective action:scalars:5}),
these logarithmic terms vanish in the limit when $H\rightarrow 0$.
Had we introduced the renormalization scale $H_0$ for the Hubble parameter
and then solved the CS equation for this case, we
could in principle get rid off all logarithmic terms.
However, as we shall see later, for the calculation
of the quantum radiative corrections to slow-roll parameters the
expression~(\ref{improved effective potential}) suffices
because our final results do not depend on $H$.
With this in mind we use 
Eq.~(\ref{improved effective potential}) in the next section.
For an alternative approach to the renormalisation group improved scalar
 effective theories in de Sitter space see 
Refs.~\cite{Elizalde:1993qh,Elizalde:1994ds}. 

\section{Slow-roll parameters}
\label{Slow-roll parameters}

 In this section we calculate the quantum\footnote{In this work when we
refer to the `classical' value of a parameter we mean
its tree-level value. When we refer to the `quantum correction' we
mean the one-loop contribution to the corresponding parameter.}
one-loop corrections to the slow-roll parameters $\epsilon $ and $\eta$
arising from the scalar matter vacuum fluctuations in inflation.
 Within the slow-roll approximation we can drop the kinetic term in
the action because it is formally second order in slow-roll
parameters~\cite{Liddle:2000cg}. That implies that -- within the
slow-roll approximation -- the leading contribution to the
stress-energy tensor is given by\footnote{In
Eq.~(\ref{stress-energy tensor}) we have neglected the tree-level
contributions to the stress-energy tensor, which are proportional
to $\xi(d/dt)^2(\varphi^2)$ and $\xi H(d/dt)(\varphi^2)$.
 These terms can be neglected based on the observation that the condition
$\dot{H}\ll H^2$ together with the slow-roll approximation imply
\begin{equation}
3\xi H\varphi\dot{\varphi}\ll \frac13 V(\varphi)\,.
\end{equation}
For a derivation of this condition we refer to
Ref.~\cite{Ante:thesis}.}
\begin{equation}
T_{\mu\nu} = -
g_{\mu\nu}\left(V_{\mathrm{RG}}(\varphi)-\frac14\frac{\delta
V_{\mathrm{RG}}(\varphi)}{\delta\ln H}\right)
\,,
\label{stress-energy tensor}
\end{equation}
where $V_{\mathrm{RG}}(\varphi)$ is the improved renormalized
effective potential~(\ref{improved effective potential}).
A straightforward calculation yields,
\begin{equation}
T_{\mu\nu} = - g_{\mu\nu}\left(\frac{\varphi^4}{4!}\, A
+3H^2\varphi^2\, B\right)\,,
\label{stress-energy tensor2}
\end{equation}
where we have introduced
\begin{eqnarray}
A & \equiv & \lambda
  + \frac{3\lambda^2}{32\pi^2}\left(X-\frac{25}{6}\right),\nonumber\\
B & \equiv & \xi
+\frac{\lambda(\xi-\frac16)}{32\pi^2}\left(X-4+\frac{1}{36(\xi-\frac16)}\right)
\,,
\label{AB}
\end{eqnarray}
and
\begin{equation}
X\equiv \ln\left(\frac{2H^2}{\lambda\varphi^2}\right) +
\psi\left(\frac12+\nu\right) + \psi\left(\frac12-\nu\right)
\,.
\label{X}
\end{equation}
The above result for the stress-energy tensor we insert into the Einstein
equation
\begin{equation}
R_{\mu\nu} -\frac12g_{\mu\nu}R = 8\pi G_N T_{\mu\nu}
\end{equation}
to obtain the following \textit{quantum Friedmann equation}
\begin{equation}
 H^2 = \frac{1}{3M_{\mathrm{Pl}}^2}\left(\frac{\varphi^4}{4!} A
+3H^2\varphi^2 B\right)
\,,
\label{QFE}
\end{equation}
where $M^2_{\mathrm{Pl}}= 1/(8\pi G_N)$. If we
take from $A$ and $B$, defined in ({\ref{AB}}), the leading
(classical) contributions, then we can from (\ref{QFE}) extract the
classical Friedmann equation in the form,
\begin{equation}
H^2_C =
\frac{\lambda\varphi^4}{72M^2_{\mathrm{Pl}}}\,\left(1-\xi\frac{\varphi^2}{M^2_{\mathrm{Pl}}}\right)^{-1}\,,
\label{CFE}
\end{equation}
which in the limit $\xi\rightarrow 0$ reduces to the well-known
result. We shall use equation (\ref{CFE}) to calculate the
number of $e$-foldings $N$ in the next section.

In order to determine the slow-roll parameters we still need
an expression for $\dot{\varphi}$.
 From Eq.~(\ref{improved effective potential}) and the slow-roll
form\footnote{Within the slow-roll approximation we can drop the
$\ddot{\varphi}$ term in Eq.~(\ref{EOM}) because that term is second
order in slow-roll parameters.} of the scalar field equation,
\begin{equation}
3H\dot{\varphi} + \frac{dV_{\mathrm{RG}}}{d\varphi} =
0\,,\label{EOM}
\end{equation}
we obtain
\begin{equation}
\dot{\varphi} = - \frac{W}{3H}
\,,
\label{varphi:dot}
\end{equation}
where
\begin{equation}
W \equiv \frac{\varphi^3}{3!}\, C + 12H^2\varphi\, D
+72\frac{H^4}{\varphi}\, E\,,
\label{W}
\end{equation}
and
\begin{eqnarray}
 C & \equiv & \lambda + \frac{3\lambda^2}{32\pi^2}\left(X-\frac{11}{3}\right) + \frac{9}{4}\frac{\lambda^3}{(4\pi)^4}\left(X-\frac{25}{6}\right)\,,\nonumber\\
 D & \equiv & \xi
 +\frac{\lambda(\xi-\frac16)}{32\pi^2}\left(X-3+\frac{1}{36(\xi-\frac16)}\right)+\frac{\lambda^2(\xi-\frac16)}{(4\pi)^4}\left(X-\frac{15}{4}+\frac{1}{48(\xi-\frac16)} \right)\,,\nonumber\\
 E & \equiv & \frac{\lambda(\xi-\frac16)^2}{(4\pi)^4}\left[\psi\left(\frac12+\nu\right) +
  \psi\left(\frac12-\nu\right)\right]\,.
\label{CDE}
\end{eqnarray}
Upon inserting the leading contributions from
Eq.~(\ref{CDE}) into Eq.~(\ref{varphi:dot}),
 the classical expression for $\dot{\varphi}$ follows immediately,
\begin{equation}
\dot{\varphi}_C =
-\frac{1}{3H_C}\left(\frac{\lambda\varphi^3}{6}+12\xi
H^2_C\varphi\right)
\,.
\end{equation}
It is important to note that with the above definitions
\begin{eqnarray}
C & = & A + \frac14\,\beta_{\lambda} + \mathcal{O}(\lambda^3)\,,\nonumber\\
D & = & B + \frac12\,\beta_{\xi} + \mathcal{O}(\lambda^2)\,.
\end{eqnarray}
We keep the parameter $E$ in (\ref{W}) for completeness, although it
yields only higher order contributions comparing to other
parameters defined by (\ref{AB}) and (\ref{CDE}).

We now proceed by making use of the standard result for the spectrum
of primordial curvature perturbation~\cite{Liddle:2000cg}

\begin{equation}
P_{\mathcal{R}}(k) = \left.\left[
\left(\frac{H}{\dot{\varphi}}\right)^2\left(\frac{H}{2\pi}\right)^2\right]\right|_{k=aH}\,.
\label{spectrum}
\end{equation}
In deriving this expression the canonical quantization of the inflaton field
has been performed in the standard way, by studying the evolution of
small perturbations around the inflaton condensate.
Since in our approach the quantum corrections are calculated
 at the level of the effective potential,
which changes the on-shell structure of the theory but does not
change the structure of Eq.~(\ref{spectrum}), we {\it conjecture}
that Eq.~(\ref{spectrum}) can be used without any further
generalization for the calculation of the one-loop quantum
corrections to the spectrum of curvature perturbation and the
implied slow-roll parameters that arise from the matter vacuum
fluctuations within the framework proposed in this work. A proof
of this conjecture is nevertheless desirable. The right-hand side
of~(\ref{spectrum}) is evaluated at the horizon exit, at which
$k=aH$, because during slow-roll inflation the Hubble parameter
does not change significantly over a few Hubble
times~\cite{Liddle:2000cg}.

The scalar spectral index $n_{s}$ is defined as
\begin{equation}
n_{s}-1 = \frac{d\ln P_{\mathcal{R}}}{d\ln k}\,, \label{spectral
index}
\end{equation}
which after some algebra, by making use of (\ref{spectrum}) and
(\ref{varphi:dot}), yields
\begin{equation}
n_s-1 =  -\frac{W}{H^4}\frac{dH^2}{d\varphi} +
\frac{2}{3H^2}\frac{dW}{d\varphi}
\,.
\label{spectral index final}
\end{equation}
By analogy with the standard result
\begin{equation}
n_s-1=-6\epsilon+2\eta\,,
\label{standard}
\end{equation}
which is valid at the classical level for various inflationary models,
we define
\begin{eqnarray}
\epsilon &\equiv & \frac{W}{6H^4}\frac{dH^2}{d\varphi}\,,\nonumber\\
\eta &\equiv & \frac{1}{3H^2}\frac{dW}{d\varphi}\,,
\label{epsiloneta}
\end{eqnarray}
such that equation (\ref{standard}) still holds for the quantum
case. On the other hand, for the gravitational wave spectrum we
use the result
\begin{equation}
P_g =
\frac{8}{M^2_{\mathrm{Pl}}}\left.\left(\frac{H}{2\pi}\right)^2\right|_{k=aH}\,.
\end{equation}
The gravitational wave spectral index $n_g$ is defined as
\begin{equation}
n_g = \frac{d\ln P_g}{d\ln k}\,, \label{gw spectral index}
\end{equation}
from which it follows that within our framework,
\begin{equation}
n_g = -\frac{W}{3H^4}\frac{dH^2}{d\varphi}\,.
\end{equation}
After taking into account the definition of $\epsilon$ from
(\ref{epsiloneta}) we obtain that the standard result,
\begin{equation}
n_g = -2\epsilon\,,
\label{ng:Q}
\end{equation}
remains valid for the quantum case as well. It is also convenient
to introduce the ratio $r$ between the gravitational wave spectrum
and the spectrum of primordial curvature perturbation,
\begin{equation}
r \equiv \frac{P_g}{P_{\mathcal{R}}}\,,
\end{equation}
which here turns into
\begin{equation}
r = \frac{8}{9M^2_{\mathrm{Pl}}}\frac{W^2}{H^4}\,.\label{ratio}
\end{equation}
With the above definitions the standard relation, $r=16\epsilon$
is not any more satisfied at the quantum
level. However, we still expect to reproduce it at the
classical limit (but see the discussion below).

The final result for $\epsilon$ and $\eta$, which includes both
the classical and quantum contributions, we write in the form
\begin{eqnarray}
\epsilon &=& \epsilon_C + \epsilon_Q\,,\nonumber\\
\eta &=& \eta_C + \eta_Q,
\end{eqnarray}
and we separate the quantum contributions into the following
two characteristic parts,
\begin{eqnarray}
\epsilon_Q &=& \frac{\beta_{\lambda}}{\lambda}\,Q_{\epsilon\lambda}
            + \beta_{\xi}\,Q_{\epsilon\xi}\,,\nonumber\\
    \eta_Q &=& \frac{\beta_{\lambda}}{\lambda}\,Q_{\eta\lambda} +
               \beta_{\xi}\,Q_{\eta\xi}
\,.
\label{convention}
\end{eqnarray}
Although the two contributions are formally of the same order of magnitude,
they have a different origin. The former contribution
in Eq.~(\ref{convention}) arises as a result of
the resummation of the mass insertions
$m_\varphi^2 =\lambda\varphi^2/2$ generated by the quartic self-interaction
in the presence of an inflaton condensate. The latter contribution
is a consequence of the resummation induced by the effective mass parameter
$12\xi H^2$ generated by the inflaton field coupled to the background
curvature. Now we shall present our results, first the classical part.

\subsection{Classical contributions $\epsilon_C$ and $\eta_C$}

After a straightforward calculation, we arrive at
\begin{eqnarray}
\epsilon_C &=&
\frac{8}{z}\frac{1-\frac12\kappa}{1-\kappa}\,,\nonumber\\
\eta_C &=& \frac{12}{z}\frac{1-\frac13\kappa}{1-\kappa}\,,
\label{classical}
\end{eqnarray}
where $z$ and $\kappa$ are defined by
\begin{eqnarray}
z &\equiv& \frac{\varphi^2}{M_{\mathrm{Pl}}^2}\,,\nonumber\\
\kappa &\equiv& \xi z = \xi\frac{\varphi^2}{M_{\mathrm{Pl}}^2} \,.
\label{zkappa}
\end{eqnarray}
It is clear that in the limit when $\xi\rightarrow 0$, i.e. when
$\kappa\rightarrow 0$, we recover the standard expressions for the
slow-roll parameters in the $\lambda\varphi^4$ inflationary model;
namely $\epsilon = 8{M_{\mathrm{Pl}}^2}/{\varphi^2}$ and
$\eta = 12{M_{\mathrm{Pl}}^2}/{\varphi^2}$. This is not surprising
since in this limit our theory reduces precisely to that inflationary model.

 We now introduce the number of $e$-foldings
\begin{equation}
N = -\int_{t_{\mathrm{end}}}^t H dt
\,,
\end{equation}
which somewhat surprisingly, when calculated classically
($H\rightarrow H_C$), gives the same result as the
${\lambda\varphi^4}$ inflationary model
\begin{equation}
N = \frac{1}{M_{\mathrm{Pl}}^2}\left(\varphi^2-
\varphi^2_{\mathrm{end}}\right)
\,.
\end{equation}
However, a mild $\xi$-dependence does enter the expression for $N$
through the value of the inflaton field at the end of inflation,
$\varphi_{\mathrm{end}}$, which is determined from the condition
$\epsilon_C(\varphi_{\mathrm{end}})=1$. From~(\ref{classical}) it
follows\footnote{The result (\ref{end}) is valid to the leading
order in $\xi$.}
\begin{equation}
\varphi^2_{\mathrm{end}} \simeq 4M^2_{\mathrm{Pl}}(2-\xi),\label{end}
\end{equation}
and finally
\begin{equation}
z\equiv \frac{\varphi^2}{M_{\mathrm{Pl}}^2} = 8\tilde{N}
\,,\qquad
\tilde{N}\simeq N+1-\frac12\xi
\,.
\label{z}
\end{equation}
We shall use the above notation when writing
the quantum contributions to slow-roll parameters.

\subsection{Quantum contributions $\epsilon_Q$ and $\eta_Q$}

In calculating the quantum contributions to slow-roll parameters we
must take into account the observational constraint required by
the near scale invariance of the spectrum
\begin{equation}
\left|\frac12\lambda\varphi^2+ 12\xi H^2 \right|\ll H^2\,.
\label{constraint}
\end{equation}
In view of equations (\ref{poles of psi})-(\ref{abc}) the
observational constraint (\ref{constraint}) implies that, in order
to study the quantum radiative corrections to slow-roll
parameters, we need the infrared limit of the RG improved theory
(\ref{improved effective potential}). This is the opposite limit
from the ultraviolet limit in which our effective theory reduces
to the Coleman-Weinberg form~(\ref{effective
action:scalars:5})~\cite{ColemanWeinberg:1973}. That means that in
order to study the quantum radiative corrections to slow-roll
parameters, one needs to focus on the infrared radiative
corrections which are specific for (quasi-)de Sitter spaces, and
completely absent in Minkowski space (which is related by a
conformal rescaling to our $n=0$ case), and hence has a very
different infrared structure. In particular, the most singular term
${a_n}/{\delta w}$ in (\ref{abc}) is absent in the conformal $n=0$
case ($a_0=0$). In summary, that means that the infrared sector plays a
crucial role in determining the quantum corrections to slow-roll
parameters.

After taking into account the constraint
(\ref{constraint}) and the expression (\ref{CFE})
for the classical Friedmann equation, we arrive at the condition
\begin{equation}
\left|\frac{9}{2\tilde{N}}-24\xi\right|\ll 1
\,,
\end{equation}
Together with the condition $8\tilde N\xi<1$, this equation then gives,
\begin{equation}
-\frac{1}{24}\left(1-\frac{9}{2\tilde N}\right)
           \ll \xi < \frac{1}{8\tilde{N}}
\,.
\label{condition for xi}
\end{equation}
Recall that typically $N$ (and hence also $\tilde{N}$)
is between 50 and 60 such that the term $9/(2\tilde N)\sim 10^{-1}\ll 1$ in
Eq.~(\ref{condition for xi}) can be to a good approximation neglected.

 We proceed by writing approximately the relation (\ref{X}) as
($\delta w\ll 1$)
\begin{equation}
X = \ln\left(\frac{\lambda\varphi^2}{2H^2}\right) -
\frac{3}{\delta w} - 2\gamma_E + \frac{7}{3} + \mathcal{O}(\delta
w )\,, \label{X2}
\end{equation}
from which it follows
\begin{equation}
\frac{dX}{d\varphi} = -\frac{1}{H^2}\frac{dH^2}{d\varphi}\,\zeta +
\frac{2}{\varphi}\,\zeta\,, \label{dXdvarphi}
\end{equation}
where we have introduced
\begin{equation}
\zeta\equiv -\left(1 + \frac{3}{\delta w
^2}\frac{\lambda\varphi^2}{2H^2}\right)\,. \label{zeta}
\end{equation}
Now the calculation of the slow-roll parameters $\epsilon$ and
$\eta$ is straightforward;
here we present only our final results. Some intermediate steps
and results can be found in Appendix A.

 For the quantum contribution~(\ref{convention}) to the slow-roll parameter
$\epsilon$ we obtain
\begin{eqnarray}
Q_{\epsilon\lambda} &=& \frac{5}{27}\,\frac{1-\frac{29}{15}\kappa
+ \frac{17}{18}\kappa^2}{(1-\kappa)^2(1-\frac{2}{3}\kappa)^2} +
\frac{1}{z}\,\frac{6-5\kappa
+\kappa^2}{1-\kappa},\nonumber\\
Q_{\epsilon\xi} &=&
-\frac{z}{18}\,\frac{\kappa}{(1-\kappa)^2(1-\frac{2}{3}\kappa)^2}
+ \frac{2}{(1-\kappa)^2}\ln\left(\frac{z}{1-\kappa}\right)
+ 4\,\frac{2-\sigma-2\kappa +
\frac{1}{2}\kappa^2}{(1-\kappa)^2}
\,,
\label{Qepsilon}
\end{eqnarray}
and for the $\eta$ slow-roll parameter defined in
(\ref{convention})
\begin{eqnarray}
Q_{\eta\lambda} &=& \frac{10}{27}\,\frac{1-\frac{29}{15}\kappa
+\frac{17}{18}\kappa^2}{(1-\kappa)^2(1-\frac{2}{3}\kappa)^2}+\frac{1}{z}\,\frac{11-10\kappa +3\kappa^2}{1-\kappa}\nonumber\\
&=&2 Q_{\epsilon\lambda}-\frac{1+\kappa}{z}\nonumber\,,
\label{Qeta}
\\
Q_{\eta\xi} &=&
-\frac{z}{9}\,\frac{\kappa}{(1-\kappa)^2(1-\frac{2}{3}\kappa)^2} + \frac{4}{(1-\kappa)^2}\,\ln\left(\frac{z}{1-\kappa}\right) + \frac{18-8\sigma-20\kappa +6\kappa^2}{(1-\kappa)^2}\nonumber\\
&=& 2 Q_{\epsilon\xi} + 2
\,.
\end{eqnarray}
In Eqs.~(\ref{Qepsilon}) and~(\ref{Qeta}) we abbreviated
\begin{eqnarray}
z &\equiv& \frac{\varphi^2}{M^2_{\mathrm{Pl}}} =
8\tilde{N}\,,\nonumber\\
\kappa &\equiv& z\xi \,,
\nonumber\\
\sigma 
      &=& \frac56 +\gamma_E + \ln 6
\,.
\label{rhosigma}
\end{eqnarray}
Since typically the number of $e$-foldings required during
inflationary epoch ranges between 50 and 60, it is evident
from~(\ref{rhosigma}) that $z$ is of the order $5\times 10^2$, which
justifies the ordering of the quantum corrections in powers of $z$.
The leading contributions are the terms linear in $z$, and they
are present only in $Q_{\epsilon\xi}$ and $Q_{\eta\xi}$ in (\ref{Qepsilon})
and (\ref{Qeta}). Both $Q_{\epsilon\xi}$ and $Q_{\eta\xi}$
contain also the next-to-leading terms of the order $\ln(z)$.
The (subleading) terms of the order $z^0$ are in fact the leading
contributions to $Q_{\epsilon\lambda}$ and
$Q_{\eta\lambda}$ in (\ref{Qepsilon}) and (\ref{Qeta}).

\subsection{Tensor and scalar spectral indices $n_g$ and $n_s$}
\label{Tensor and scalar spectral indices $n_g$ and $n_s$}

 We can now easily calculate the tensor and scalar spectral indices from
the results for the slow-roll parameters
 $\epsilon$ and $\eta$~(\ref{classical}), (\ref{Qepsilon}--\ref{Qeta}).

 Note first that the tensor spectral index $n_g$~(\ref{ng:Q})
can be trivially obtained by summing the classical~(\ref{classical})
and quantum~(\ref{Qepsilon}) contributions for $\epsilon$, since from our
definition~(\ref{epsiloneta}) it follows that Eq.~(\ref{ng:Q})
is valid also at the quantum level.

 Next we consider the scalar spectral index $n_s$.
By making use of Eqs.~(\ref{classical}) and~(\ref{Qepsilon}--\ref{Qeta}) and
separating again the classical and quantum contributions as,
\begin{equation}
n_s -1 = (n_s-1)_{C} + (n_s-1)_Q
\,,
\label{nsQ}
\end{equation}
where
\begin{equation}
(n_s-1)_Q = \frac{\beta_{\lambda}}{\lambda}\,Q_{(n_s-1)_{\lambda}}
+ \beta_{\xi}\,Q_{(n_s-1)_{\xi}}\,,
\end{equation}
we arrive at the classical scalar spectral index,
\begin{equation}
(n_s-1)_{C} =
-\frac{24}{z}\,\frac{1-\frac23\kappa}{1-\kappa}
\,,\qquad
  z = 8\tilde N\,,\qquad \kappa = 8\tilde N\xi
\,.
\label{nsC}
\end{equation}
The quantum contributions are given by,
\begin{eqnarray}
Q_{(n_s-1)_{\lambda}} &=&
 -\frac{10}{27}\,
\frac{1-\frac{29}{15}\kappa+\frac{17}{18}\kappa^2}
     {(1-\kappa)^2(1-\frac{2}{3}\kappa)^2}
-\frac{1}{z}\,\frac{14-10\kappa}{1-\kappa}
\,,
\label{nsQ2}
\\
Q_{(n_s-1)_{\xi}} &=&
\frac{z}{9}\,\frac{\kappa}{(1-\kappa)^2(1-\frac{2}{3}\kappa)^2} - \frac{4}{(1-\kappa)^2}\,\ln\left(\frac{z}{1-\kappa}\right) - \frac{12-8\sigma-8\kappa}{(1-\kappa)^2} \,.
\nonumber
\end{eqnarray}
Notice that the quantum contributions
$({\beta_{\lambda}}/{\lambda})Q_{(n_s-1)_{\lambda}}$ and
$\beta_{\xi}Q_{(n_s-1)_{\xi}}$ are both much smaller than the
classical contribution $(n_s-1)_C$ due to the fact that both
${\beta_{\lambda}}/{\lambda}$ and $\beta_{\xi}$ are of the order of
$\lambda$, which is constrained by experimental data to be of the
order of $\lambda \sim 10^{-12}$ (we provide a more precise
constraint for $\lambda$ below). The quantum contributions can
become significant only in an inflationary model in which the
relevant coupling constant can be as large as the order of $10^{-3}$.
It is important to notice that $Q_{(n_s-1)_{\lambda}}$ provides an
infrared enhanced quantum correction proportional to the number of
$e$-foldings $N$, while, on the other hand, $Q_{(n_s-1)_{\xi}}$
contains a correction which is enhanced by $N^2$ when compared to
the na\"ive expectation ${\cal O}(1/z)\sim 1/N)$. 

This shows that in the de Sitter
invariant limit the quantum corrections to slow-roll parameters
accumulate only from the time the mode becomes super-Hubble until
the end of inflation. For a mode which exits horizon $N$
$e$-foldings before the end of inflation, the whole history of
inflation before the Hubble exit ({\it i.e.} when the mode was
sub-Hubble) is completely irrelevant and does not contribute in a
cumulative manner to the quantum corrections. This disagrees with
the result found in Refs.~\cite{Sloth:2006az,Sloth:2006nu}, where
it is claimed that the quantum loop corrections induce corrections
to the slow-roll parameters and spectral indices which depend on
the total duration of inflation and are thus enhanced when
inflation lasts for a large number of $e$-foldings.
 This $N^2$ enhancement is the main result of our work, and it
resolves the Weinberg's dilemma \cite{Weinberg:2005,Weinberg:2006}:
how big can be the correction induced by the quantum fluctuations
of light or massless scalar fields during inflation,
given the fact that the (equal time and space) scalar
field correlator for a massless minimally coupled scalar grows
linearly with time during (de Sitter) inflation,
\begin{equation}
\left<0\right|\varphi^2(x)\left|0\right> = \mathrm{infinite}
+\frac{H^2}{4\pi^2}\ln a\,,
\end{equation}
or
\begin{equation}
\left.\left(\left<0\right|\varphi^2(x)\left|0\right>\right)\right|_{\mathrm{fin}}
\simeq \frac{H^2}{4\pi^2}N\,,
\end{equation}
where here $N$ denotes (minus) the number of $e$-foldings.

 By comparing Eqs.~(\ref{ng:Q}) (\ref{Qepsilon}) and (\ref{nsQ2}) 
we observe that the following curious relation holds, 
\begin{equation}
(n_g)_Q = (n_s-1)_Q 
        + \frac{\beta_\lambda}{\lambda}
                \left(\frac{2(1+\kappa)}{z}
                \right)
         -4 \beta_\xi
\,,
\label{ng=ns-1:Q}
\end{equation}
such that the leading quantum contributions ${\cal O}(\lambda \tilde N)$
and ${\cal O}(\lambda \ln(\tilde N))$
to the tensor and scalar spectral indices are equal. 
This approximate equality can be traced back to the fact that 
$\epsilon_Q\simeq \eta_Q/2$, from where it follows that 
$[(W/H^4)(dH^2/d\varphi)]_Q\simeq [(1/H^2) (dW/d\varphi)]_Q$,
or equivalently $[d(W/H^2)/d\varphi]_Q\simeq 0$. The expression 
$W/H^2$ is proportional to the square-root of the ratio 
$r=P_g/P_{\cal R}$, which according to Eq.~(\ref{rfinal}) 
does not receive any quantum corrections that are amplified by $N$. That means
that, because the tensor and scalar spectra are identically 
affected by the quantum corrections enhanced by $N$, 
these leading corrections cancel in the ratio $r$.
We do not have a deeper insight to why that is the case. 
Note that the approximate equality~(\ref{ng=ns-1:Q}) does not 
hold for the corresponding classical parts~(\ref{classical}) and~(\ref{nsC}).
 If inflationary models with large quantum corrections are found, 
Eq.~(\ref{ng=ns-1:Q}) could be used to resolve the quantum
from classical contributions to the tensor and scalar spectral indices.

\subsection{The spectrum of curvature perturbation ${\cal P}_{\cal R}$
          and the tensor-to-scalar ratio $r$}

 Next we consider the spectrum of curvature perturbation~(\ref{spectrum}).
As usual we decompose the spectrum into the classical and quantum parts as,
\begin{equation}
P_\mathcal{R} = \left(P_\mathcal{R}\right)_C +
\left(P_\mathcal{R}\right)_Q
\,,
\label{PR:+}
\end{equation}
where
\begin{equation}
\left(P_\mathcal{R}\right)_Q \equiv
\frac{\beta_\lambda}{\lambda}Q_{P_\mathcal{R}\lambda} + \beta_\xi
Q_{P_\mathcal{R}\xi}
\,.
\end{equation}
Working within our framework we obtain,
\begin{equation}
\left(P_\mathcal{R}\right)_C =
\frac{\lambda}{9\pi^2}\frac{\tilde{N}^3}{1-\kappa}\,,
\label{P_R:C}
\end{equation}
and
\begin{eqnarray}
\frac{Q_{P_\mathcal{R}\lambda}}{\left(P_\mathcal{R}\right)_C} &=&
- \frac{z}{27}\frac{1-\frac{25}{24}\kappa}{(1-\kappa)(1-\frac23\kappa)} +
\frac12\ln\left(\frac{z}{1-\kappa}\right) - \frac12(1-\kappa)
- \sigma - \frac1{12}
\nonumber\,,
\\
\frac{Q_{P_\mathcal{R}\xi}}{\left(P_\mathcal{R}\right)_C} &=&
-\frac{z^2}{24}\frac{1}{(1-\kappa)(1-\frac23\kappa)} +
\frac{z}{2}\frac{1}{1-\kappa}\ln\left(\frac{z}{1-\kappa}\right) -
z\left(\frac{\sigma}{1-\kappa}+1\right)
\,,
\label{PR:Q}
\end{eqnarray}
where $\tilde{N}=N+1-\frac12\xi$, $z=8\tilde N$ and
$\kappa = 8\xi\tilde N$. This implies that, just like the spectral indices,
when compared to the classical contribution, the quantum contribution
to the spectrum is suppressed as $\lambda \tilde N^2$, and thus
unobservably small for the model in consideration.

 The spectrum of curvature perturbation is an observable quantity and
the three year WMAP data provide a strong
constrain for it~\cite{Spergel:2006},
\begin{equation}
P_{\mathcal{R}} \approx 29.5\times 10^{-10}A\,,
\label{P_R:obs}
\qquad
%
%
A = 0.801^{+0.043}_{-0.054}
\,.
\end{equation}
In the case of a weak coupling to the background,
{\it i.e.} when $8|\xi|\tilde{N}\ll 1$ and $\tilde{N}\approx N+1$,
Eqs.~(\ref{P_R:C}) and (\ref{P_R:obs}) imply
\begin{eqnarray}
\lambda_{50} \approx 1.58\times 10^{-12}\,,\qquad
\lambda_{60} \approx 9.25\times 10^{-13}\,,
\qquad (8|\xi|\tilde{N}\ll 1)
\,,
\end{eqnarray}
where the subscripts on $\lambda$ denote the number of $e$-foldings $N$.
We stress that these values for $\lambda$ are strictly speaking valid only
for this particular model. Indeed, when we choose the coupling to the
background in the range,
\begin{equation}
\frac{1}{24}\gg - \xi > \frac{1}{8\tilde{N}}
\,,
\label{condition:final:2}
\end{equation}
then from Eq.~(\ref{P_R:C}) we see that the value of
$\lambda$ can be up to one order of magnitude larger,
\begin{equation}
  \lambda_{50}\simeq 2.69\times 10^{-11}\times (-24 \xi)
\,,\qquad
  \lambda_{60}\simeq 1.88\times 10^{-11}\times (-24 \xi)
\,,\qquad
(-24 \xi\ll 1)
\,.
\label{lambda50+60:2}
\end{equation}
These larger values of $\lambda$ are still too small
however to render the quantum effects observable.

In other inflationary models the relation for
the spectrum of curvature perturbation
(\ref{P_R:C}) can in general be different, hence allowing for models in
which couplings are larger. Furthermore, the quantum effects in some other
models may be much stronger -- of particular interest are hybrid
inflationary models~\cite{ProkopecBilandzic}.

 Let us now consider the ratio $r$ of the gravitational wave and
curvature spectrum. From Eq.~(\ref{ratio}) we obtain
\begin{eqnarray}
r &=& r_C + r_Q
\nonumber\\
r_C &=& \frac{128}{z}
\label{r:C}
\\
r_Q &=& \frac{\beta_{\lambda}}{\lambda}\,\frac{64}{z}\,(1-\kappa) +
       128\beta_{\xi}
\,,
\label{rfinal}
\end{eqnarray}
where again $z\equiv 8\tilde{N}$ and  $\kappa\equiv z\xi$. We see
that $r_C = {128}/{z}$, and after observing from
(\ref{classical}) that $\epsilon_C = {8}/{z}$ in the limit
when $\xi\rightarrow 0$ (i.e. when $\kappa\rightarrow 0$), we
obtain the standard result $r=16\epsilon$.
This relation is violated both by the classical corrections
from  $\kappa\neq 0$ ($\xi\neq 0$) and by the quantum contributions.
In particular, when $\kappa= 8\tilde{N}\xi \ll -2$, one gets
$r\simeq 32\epsilon$. When compared with the classical
contribution~(\ref{r:C}), the quantum contribution to
 $r$~(\ref{rfinal}) is as usual suppressed
by $\lambda$, but --  in contrast to the slow-roll
 parameters $\epsilon_Q$ and $\eta_Q$ --
 it is not enhanced by powers of $N$.

 It is well known that the minimally coupled
${\lambda\varphi^4}$ inflationary model is disfavored
by observations~\cite{Spergel:2006} by about 2 standard deviations.
This is not however in general the case with the nonminimally coupled
 ${\lambda\varphi^4}$ inflationary model. Indeed,
 from Eqs.~(\ref{condition for xi}) and~(\ref{nsC})
we infer that in the range
\begin{equation}
\frac{1}{24}\gg - \xi > \frac{1}{8\tilde{N}}
\,.
\label{condition:final}
\end{equation}
the deviation of the classical spectral index of scalar curvature
perturbation~(\ref{nsC}) from scale invariance
is reduced approximately by a factor of 2/3,
\begin{equation}
(n_s-1)_{C} \approx -\frac{16}{z} = - \frac{2}{\tilde N}
\,,\qquad
  (-\xi \gg (8\tilde{N})^{-1})
\,,
\label{nsC:2}
\end{equation}
while the value of $r_C$~(\ref{r:C}) remains unchanged. This then
implies that, similarly as in the minimally coupled massive
inflationary model, the $\lambda\varphi^4$ model with $\xi$ in the
range~(\ref{condition:final}) falls roughly at the $1\sigma$
contour of Figure 14 in Ref.~\cite{Spergel:2006}, rendering these
nonminimally coupled $\lambda\varphi^4$ inflationary models
consistent with the three year WMAP
data~\cite{Spergel:2006,Tsujikawa:2004my,KomatsuFutamase:1999,FakirUnruh:1990}.
We emphasize that, because $\xi=\xi(\varphi_0)$ runs logaritmically
towards its infrared fixed point $\xi_{\rm FP} = 1/6$, it is
natural to assume that $\xi$ deviates from zero. Indeed, even if
we choose $\xi=0$ (which corresponds to the coupling at some scale
$\varphi=\varphi_0$), the running of $\xi$ will induce the
dominant quantum contributions to slow-roll parameters. Thus for
consistency it is necessary to consider the effects of nonminimal
coupling, and choosing $\xi$ in the range~(\ref{condition:final})
is {\it a priory}
 as natural as any other choice (different, of course, from $\xi=1/6$).

In conclusion, we have found out that, even though the quantum
effects to slow-roll parameters are enhanced by the number of
$e$-foldings squared, they are suppressed by the small coupling
constant $\lambda$. Due to the smallness of $\lambda$ however, the
quantum effects have a negligible impact on the plot presented for
example in Figure 14 of \cite{Spergel:2006}.

\section{Discussion}
\label{Discussion}

In this work we develop a quantum field theoretic framework within which
the quantum corrections to slow-roll parameters and
observables from inflationary models can be calculated.
Whe main purpose of this paper is methodical, and we postpone a detailed study
of the quantum corrections to inflationary observables in various
inflationary models to a future work~\cite{ProkopecBilandzic}.

 We illustrate how our framework works by performing the relevant calculations
in a concrete inflationary model chosen for its simplicity.
More specifically, we consider a $\lambda\varphi^4$ inflationary
model~(\ref{scalar tree action}--\ref{scalar:potential})
with a nonminimal coupling to the background curvature.
Our formalism can be quite straightforwardly generalized to other
inflationary models. Within our
$\lambda\varphi^4$ model we calculate the quantum corrections to
the inflationary slow-roll parameters $\epsilon$ and
$\eta$~(\ref{convention}), (\ref{Qepsilon}--\ref{Qeta}),
based on which we derive the scalar spetral index
$n_s$~(\ref{nsQ}--\ref{nsQ2}),
the tensor spectral index $n_g$~(\ref{ng:Q}), (\ref{Qeta})
and the spectrum of curvature perturbation
${P}_{\cal R}$~(\ref{PR:+}--\ref{PR:Q}).
These corrections arise from the one-loop scalar vacuum fluctuations
during de Sitter inflation.
We find that the dominant quantum effects
for the spectral indices~(\ref{ns-1:Q}), (\ref{ng:Q}), (\ref{Qeta})
are suppressed as $\lambda N^2$
when compared to the classical (tree-level) results~(\ref{ns-1:C}),
(\ref{classical}). The dominant quantum contribution
arises from the inflaton coupling to the background curvature.

 Our theoretic framework can be improved in several aspects.
For example, one could generalize our calculation of the renormalization
group improved effective action~(\ref{improved effective potential})
to quasi-de Sitter spaces, which are important since these spaces
comprise a large fraction of inflationary models.
Next one should generalize our analysis to incorporate
other matter degrees of freedom which would allow us to
incorporate a broad spectrum of inflationary models.
Even more importantly one should study the role
of the interactions that couple matter and gravitational degrees of freedom.


\section*{Appendix A}
Here we provide a more detailed derivation for the quantum
contributions to slow-roll parameters given by the relations
(\ref{Qepsilon}) and (\ref{Qeta}). We begin by abbreviating
the equalities in Eq.~(\ref{AB}) as
\begin{eqnarray}
A &=& \lambda + \beta_{\lambda}X_A\,,\nonumber\\
B &=& \xi + \beta_{\xi}X_B\,,
\end{eqnarray}
where now
\begin{eqnarray}
X_A &\equiv& \frac12\left(X-\frac{25}{6}\right)\,,\nonumber\\
X_B &\equiv&
\frac12\left(X-4+\frac{1}{36\left(\xi-\frac16\right)}\right)\,.
\end{eqnarray}
It is important to note that, in order to strictly follow our
notation in Eq.~(\ref{convention}), the last term in the definition of
$X_B$ actually contributes to both $Q_{\epsilon\lambda}$ and
$Q_{\eta\lambda}$. The reason is that, after taking into account
the relations (\ref{betafunctionsfinal}) for $\beta_{\lambda}$ and
$\beta_{\xi}$, it follows immediately that,
\begin{equation}
\beta_{\xi}\frac{1}{72\left(\xi-\frac16\right)} =
\frac{1}{72}\frac{\lambda}{16\pi^2} =
\frac{1}{216}\,\frac{\beta_{\lambda}}{\lambda}\,.
\end{equation}
The origin of this mixing is the peculiar ${1}/{36}$ term in
(\ref{improved effective potential}), which is suppressed by
$\lambda$, but not by $\xi -\frac16$.

It is convenient to introduce the parameter $\alpha$ as follows,
\begin{equation}
\alpha \equiv \left(\frac{M_{\mathrm{Pl}}^2}{\varphi^2}-
\xi\right)^{-1}\,,
\end{equation}
which, from the definitions in (\ref{zkappa}), can also be
expressed as
\begin{equation}
\alpha = \frac{z}{1-\kappa}\,.\label{alpha}
\end{equation}
Assuming that $\xi{\varphi^2}/{M_{\mathrm{Pl}}^2}<1$ is far
enough from $1$, from the quantum Friedmann
equation (\ref{QFE}) we obtain the following expression,
\begin{equation}
\frac{\varphi^2}{H^2} \simeq
\frac{72}{\alpha\lambda}\left(1-\beta_{\xi}\alpha X_B -
\frac{\beta_{\lambda}}{\lambda}X_A\right)\,.\label{QFE2}
\end{equation}
Upon differentiating the quantum Friedmann
equation (\ref{QFE}) with respect to $\varphi$ we get,
\begin{equation}
\frac{dH^2}{d\varphi} = \frac{\alpha\varphi}{18}\,\frac{\lambda +
\beta_{\lambda}\zeta_A +
36\frac{H^2}{\varphi^2}\left(\xi+\beta_{\xi}\zeta_B\right)}{1+\beta_{\lambda}\frac{\varphi^2}{H^2}\frac{\zeta\alpha}{144}+
\beta_{\xi}\alpha\left(\frac{\zeta}{2}-X_B\right)}\label{dH^2dvarphi}\,.
\end{equation}
In deriving this expression we have used,
\begin{eqnarray}
\frac{dA}{d\varphi} &\simeq& \frac{1}{\varphi}\beta_{\lambda}
+\frac{1}{2}\,\frac{dX}{d\varphi}\beta_{\lambda}\,,\nonumber\\
\frac{dB}{d\varphi} &\simeq& \frac{1}{\varphi}\beta_{\xi}
+\frac{1}{2}\,\frac{dX}{d\varphi}\beta_{\xi}\,,
\end{eqnarray}
where ${dX}/{d\varphi}$ and $\zeta$ are given by Eqs.~(\ref{dXdvarphi})
and~(\ref{zeta}), respectively. In writing
Eq.~(\ref{dH^2dvarphi}) we have also introduced
\begin{eqnarray}
\zeta_A &\equiv& X_A+ \frac14(1+\zeta)\,,
\nonumber\\
\zeta_B &\equiv& X_B+\frac12(1+\zeta)\,.
\end{eqnarray}
In order to evaluate $\epsilon$ from (\ref{epsiloneta}) we still need the
expression for $W$. With the above definitions, from (\ref{W}) it
follows
\begin{equation}
W \simeq \frac{\lambda\varphi^3}{6}\left[1+\alpha\xi
+\frac{\beta_{\lambda}}{\lambda}\left(\frac14 +X_A(1+\alpha\xi)\right)
 +\beta_{\xi}\alpha\left(\frac12+X_B(1+\alpha\xi)\right)\right]
\,.
\label{W2}
\end{equation}
What remains to be done is to expand the denominator in
(\ref{dH^2dvarphi}) to the linear order in $\beta_{\lambda}$ and
$\beta_{\xi}$ and then insert the resulting expression together
with (\ref{QFE2}) and (\ref{W2}) into the definition of $\epsilon$ given in
(\ref{epsiloneta}). In this manner both the classical
(\ref{classical}) and the quantum contributions ({\ref{Qepsilon}}) can
be obtained.

To calculate $\eta$, we must determine ${dW}/{d\varphi}$.
After some algebra we arrive at
\begin{eqnarray}
\frac{dW}{d\varphi}&\simeq&\frac{dH^2}{d\varphi}\,12\varphi\,\left[\xi
-\frac{\beta_{\lambda}}{\lambda}\frac{\zeta}{2\alpha}
+ \beta_{\xi}\left((X_B+\frac12)-\frac12\,\zeta\right)\right]\nonumber\\
&&+\,\frac12\,\varphi^2\left[\lambda+\beta_{\lambda}\left(X_A +
\frac{\zeta}{3}+\frac{7}{12} \right)\right]\nonumber\\
&&
+\,12H^2\left[\xi+\beta_{\xi}\left(X_B+\zeta+\frac32\right)\right]\,,\label{dWdvarphi}
\end{eqnarray}
where ${dH^2}/{d\varphi}$ is given in (\ref{dH^2dvarphi}).
After expanding the denominator in (\ref{dH^2dvarphi}) and after
inserting (\ref{QFE2}) and (\ref{dWdvarphi}) into the definition
for $\eta$ given by (\ref{epsiloneta}), both the classical
(\ref{classical}) and quantum contribution (\ref{Qeta}) are obtained.

At the end, we summarize
\begin{eqnarray}
X_A &\simeq& \frac12\left(\ln\alpha -\frac{3}{\delta\omega}\right)
-\sigma-\frac{1}{12}
\,,\nonumber\\
X_B &\simeq& \frac12\left(\ln\alpha
-\frac{3}{\delta\omega}+\frac{1}{36\left(\xi-\frac16\right)}\right)-\sigma\,,\nonumber\\
\delta w &\simeq& \frac{36}{\alpha} + 12\xi\,,\nonumber\\
\zeta &\simeq&
-\left[1+\frac34\,\frac{\alpha}{(3+\alpha\xi)^2}\right]\,,
\end{eqnarray}
where $\alpha$ is determined by (\ref{alpha}), while
$\sigma$ is given by (\ref{rhosigma}).

\end{document}